\documentclass[twocolumn,showpacs,preprintnumbers,amsmath,amssymb,prl]{revtex4}
\usepackage{graphicx}
\usepackage{dcolumn}
\usepackage{bm}
\usepackage[usenames,dvipsnames]{color}
\usepackage{ulem}

\begin{document}

\title{Peierls Mechanism of the Metal-Insulator Transition in \\
Ferromagnetic Hollandite K$_2$Cr$_8$O$_{16}$}

\author{T. Toriyama,$^1$ A. Nakao,$^2$ Y. Yamaki,$^3$ H. Nakao,$^2$ Y. Murakami,$^2$ 
K. Hasegawa,$^4$ M. Isobe,$^4$ Y. Ueda,$^4$ A. V. Ushakov,$^5$ D. I. Khomskii,$^5$ 
S. V. Streltsov,$^{6,7}$ T. Konishi,$^8$ and Y. Ohta$^1$}

\affiliation{$^1$Department of Physics, Chiba University, Chiba 263-8522, Japan}
\affiliation{$^2$Condensed Matter Research Center and Photon Factory, IMSS, KEK, Tsukuba 305-080, Japan}
\affiliation{$^3$Department of Physics, Tohoku University, Sendai 980-8576, Japan}
\affiliation{$^4$Institute for Solid State Physics, University of Tokyo, Kashiwa 277-8581, Japan}
\affiliation{$^5$II. Physikalisches Institut, Universit\"at zu K\"oln, Z\"ulpicher Stra\ss e 77, D-50937 K\"oln, Germany}
\affiliation{$^6$Institute of Metal Physics, S. Kovalevskoy St.~18, 620041 Ekaterinburg GSP-170, Russia}
\affiliation{$^7$Ural Federal University, Mira St.~19, 620002 Ekaterinburg, Russia}
\affiliation{$^8$Graduate School of Advanced Integration Science, Chiba University, Chiba 263-8522, Japan}

\date{28 December 2010}

\begin{abstract}
Synchrotron X-ray diffraction experiment shows that the 
metal-insulator transition occurring in a ferromagnetic 
state of a hollandite K$_2$Cr$_8$O$_{16}$ is accompanied 
by a structural distortion from the tetragonal $I4/m$ 
to monoclinic $P112_{1}/a$ phase with a 
$\sqrt{2}\times\sqrt{2}\times 1$ supercell.  
Detailed electronic structure calculations demonstrate 
that the metal-insulator transition is 
caused by a Peierls instability in the quasi-one-dimensional 
column structure made of four coupled Cr-O chains running in the 
$c$-direction, leading to the formation of tetramers of Cr ions 
below the transition temperature.  
This furnishes a rare example of the Peierls transition of 
fully spin-polarized electron systems.  
\end{abstract}


\maketitle

Low-dimensional solids have attracted considerable attention in 
recent years.  An interesting class of these materials, which 
is just started to be explored, are the so-called tunnel compounds.  
Typically, they are made of double $M$O chains ($M$ is, e.g., a 
transition-metal element) with edge-sharing $M$O$_6$ octahedra, 
forming zigzag chains (triangular ladders).  
These double chains are connected {\it via} corner-sharing 
octahedra into particular structures, e.g., hollandites, 
ramsdellites, romanechites, todorokites, or systems with the 
calcium-ferrite structure.  
The large tunnels thus formed are filled by ions such as 
K$^{+}$, Rb$^{+}$, Ca$^{2+}$, Ba$^{2+}$, etc.  
These materials display a wide range of magnetic, charge, and 
orbital orderings, metal-insulator transitions (MIT), etc 
\cite{isobe,hasegawa,yamaura,pieper}.  
They are usually considered as quasi-one-dimensional (1D), 
the 1D building blocks being the double chains \cite{pieper,chern}.  

Perhaps the most surprising phenomenon was discovered recently 
in a Cr hollandite K$_2$Cr$_8$O$_{16}$ \cite{hasegawa}.  This 
material, which is tetragonal (space group $I4/m$) \cite{tamada} 
and metallic at room temperature, becomes ferromagnetic below 
$T_c=180$ K, but at lower temperatures, it experiences an MIT 
at $T_{\rm MI}=95$ K, remaining ferromagnetic in the insulating 
phase.  This is very unusual: typically, insulating transition-metal 
oxides are antiferromagnetic, and ferromagnetism usually goes hand 
in hand with metallicity \cite{khomskii97}.  
Among the several explanations offered, a proposal was made that 
charge ordering is responsible for the low-temperature behavior 
\cite{mahadevan}.  

The clue to understanding the nature of MIT in K$_2$Cr$_8$O$_{16}$ 
can be obtained from structural studies, which should show the 
signatures of eventual charge or orbital order, or some other 
structural distortions.  Conventional powder X-ray studies 
\cite{hasegawa} did not detect any structural distortions, which 
nevertheless should be present.  Therefore, in the present study, 
we undertook careful structural determination, using synchrotron 
radiation, and we combined it with the detailed electronic 
structure calculations to obtain a complete picture of the 
phenomena occurring in K$_2$Cr$_8$O$_{16}$ at the MIT.  

\begin{figure}[htbp]
\begin{center}
\includegraphics[scale=0.41]{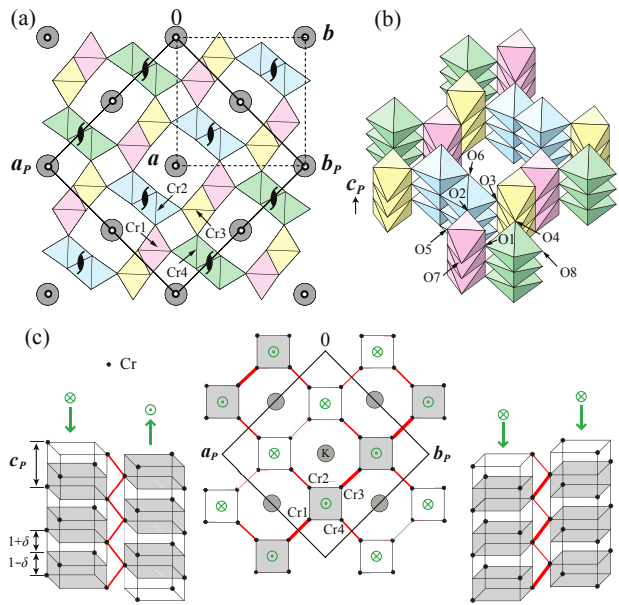}
\caption{(Color online) (a) Crystal structure of K$_2$Cr$_8$O$_{16}$ 
at 20 K viewed down the $c$-axis and (b) perspective view showing 
the connection of the double Cr chains running in the $c$-direction.  
The structures are drawn with the CrO$_6$ octahedra.  
In (a), the unit cells of the FM and FI phases are indicated by 
dotted and solid lines, respectively, and the inversion centers and 
two-fold screw axes are indicated by open circles and solid symbols, 
respectively.  
(c) Schematic illustration of the FI phase, emphasizing the four-chain 
columns with the lattice dimerization, as indicated by shadow.  
Red lines mark double zigzag chains.  The four-chain columns with 
$\uparrow$/$\downarrow$ dimerization (indicated by $\odot$/$\otimes$) 
are arranged in a stripe-like pattern.}
\label{fig1}
\end{center}
\end{figure}

The picture which emerged out of our study is rather unexpected: 
one can indeed consider this material as quasi-1D, but the 1D 
building blocks are not the usually considered double chains, but 
rather the four corner-sharing chains around small squares, forming 
vertical ``chimneys'' in the $c$-direction (columns Cr1-Cr2-Cr3-Cr4 
in Fig.~\ref{fig1}).  
To the best of our knowledge, this is the first example of the 
1D structure made in such a way.   
The MIT in K$_2$Cr$_8$O$_{16}$ is caused by the Peierls instability 
in this four-chain column.  This mechanism provides a natural 
explanation for the observed structural distortions, i.e., the 
lattice dimerization in each of the four-chain columns below 
$T_{\rm MI}$.  
Importantly, we detected no charge ordering at the MIT; charge 
density is found to be the same at all Cr ions.  

The synchrotron X-ray diffraction experiment was carried out on the 
imaging-plate  diffractometer at beamlines 8A and 8B of Photon 
Factory in KEK.  The reflection data sets for the structural analysis 
on single crystals were collected in the ferromagnetic metallic (FM) 
and ferromagnetic insulating (FI) phases using the X-ray energy of 
18~keV \cite{SM}.  

To elucidate the origin of the FI phase, the crystal structure was 
explored by oscillation photographs.  Superstructure reflections 
associated with the unit cell of $\sqrt{2}a\times\sqrt{2}b\times c$ 
were clearly observed in the FI phase (20~K).  
The intensity of the superlattice 
reflections is $\sim$10$^{-3}$ of the fundamentals, continuously 
increasing with decreasing temperature, suggesting a second-order phase 
transition.  The distribution and intensity of the superlattice 
reflections are consistent with the Laue class 2/$m$; although peak 
splittings due to distortion of the lattice from 90 degree were not 
observed, the intensities of equivalent/inequivalent superlattice 
reflections showed $I(hkl)=I(\overline{h}\overline{k}l)\neq 
I(\overline{k}hl)=I(\overline{k}h\overline{l})$ with the monoclinic 
domain ratio of $6:4$.  Thus, the system is monoclinic in the FI phase.  
From a detailed analysis of the reflection conditions \cite{SM}, 
we concluded that the space group of the FI phase is $P112_1/a$.  
The transition is accompanied by a symmetry breaking with the loss 
of the four-fold axis and the mirror plane perpendicular to the $c$-axis.  

The structure analysis of the FI phase was performed on the basis of 
the space group $P$112$_{1}$/$a$, where four Cr sites, two K sites, 
and eight O sites become crystallographically inequivalent \cite{SM}.  
From the result, we nevertheless found that the Cr valences estimated 
from the bond valence sum are the same at all the Cr sites within 
the experimental error ($\lesssim$0.03 electrons/Cr); i.e., no 
evidence of charge ordering.  The resulting crystal structure of 
the FI phase is shown in Fig.~\ref{fig1}.  

The observed structural distortions can be explained as follows:  
The four Cr sites form three types of double chains (Cr1-Cr3, Cr2-Cr2, 
and Cr4-Cr4 double chains, see Fig.~\ref{fig1}(a)), a half of which 
(Cr1-Cr3) has a Cr-Cr bond alternation.  Simultaneously, there occurs 
a lattice dimerization in all the four-chain columns made of 
Cr1-Cr2-Cr3-Cr4 chains, leading to the formation of tetramers of Cr 
ions (shown in Fig. 1(c) by shading).  This dimerization takes place 
within the unit cell in the $c$-direction, without doubling the $c$ 
lattice parameter.  The degree of dimerization is $\delta=0.025$ at 
20 K, where $\delta$ is defined in Fig.~\ref{fig1} and is zero in 
the FM phase.  As we show below, it is this lattice dimerization 
in the four-chain columns which plays a crucial role and drives the 
structural and metal-insulator transition.  
The relative stacking of column distortions is in such a way that 
they form stripes in the ${\bm a}_P$ direction (see Fig.~\ref{fig1}(c), 
middle), with the distortions in-phase along these stripes, but 
opposite (shifted by a half of the period) in the neighboring stripes.  
This leads to the net monoclinic symmetry.  
This also explains the occurrence of the bond alternation in the 
double chains made of Cr1-Cr3 in these stripes (see Fig.~\ref{fig1}(c), 
right), but no dimerization in double chains made of Cr2-Cr2 and 
Cr4-Cr4 which are sandwiched between the stripes with opposite 
distortions (Fig.~\ref{fig1}(c), left).  

\begin{figure}[thbp]
\begin{center}
\includegraphics[scale=0.58]{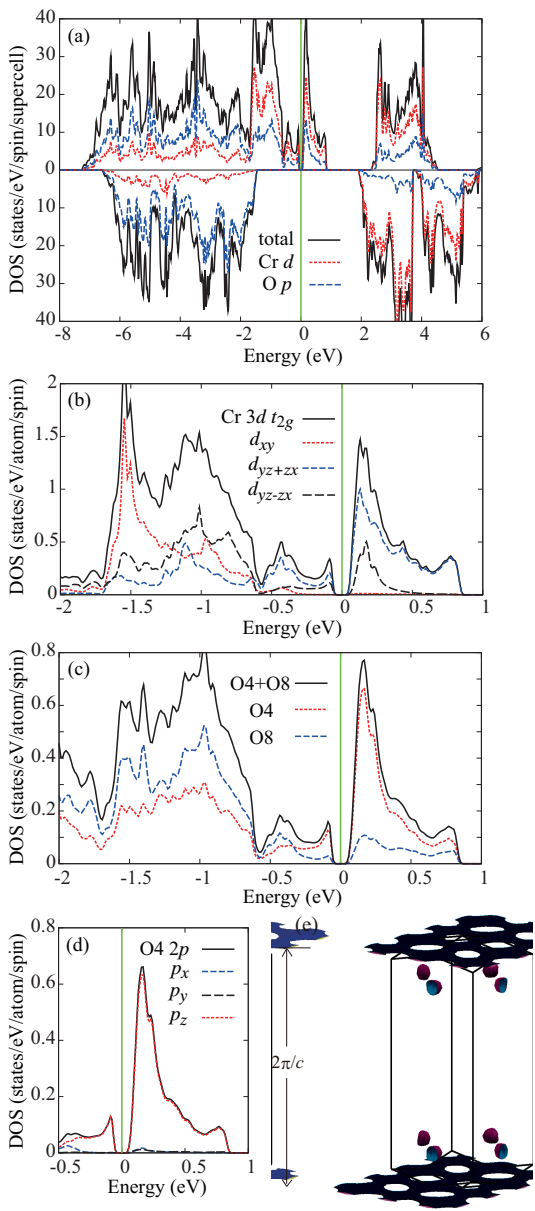}
\caption{(Color online) Calculated GGA+$U$ partial DOS in the FI 
phase of K$_2$Cr$_8$O$_{16}$ at $U=3$ eV: 
(a) the majority- and minority-spin DOS in a wide energy range, 
(b) the majority-spin $3d$ $t_{2g}$ orbitals of Cr4, 
(c) the majority-spin $2p$ orbitals of corner (O4) and 
edge (O8) oxygens, and 
(d) the orbital decomposition for the majority-spin band of O4.  
The Fermi level is indicated by a vertical line.  
In (e), the calculated Fermi surface in the FM phase at 
$U=2.9$ eV is shown.}  
\label{fig2}
\end{center}
\end{figure}

To clarify the mechanism of the MIT further, we carried out the 
electronic structure calculations using different methods: 
local-density approximation (LDA) and generalized gradient approximation 
(GGA), taking into account the Hubbard-type repulsion (LDA/GGA$+U$), 
in the FLAPW \cite{wien2k} and LMTO-ASA \cite{LMTO} implementations.  
The double counting corrections were taken in form of the fully 
localized limit \cite{czyzyk} both for the FLAPW and LMTO-ASA 
calculations.  We also checked the around-mean-field correction 
\cite{czyzyk} in the FLAPW calculation and confirmed that the 
results are almost the same.  
We also note that the value $U\simeq 2-3$ eV we used is standard 
and accepted for Cr oxides with Cr$^{3+}$ and Cr$^{4+}$ 
\cite{mazin,korotin}; we actually confirmed that the results do 
not change qualitatively for different values of $U$ between 2 
and 5 eV.  

We then found that the electronic structure of the high-temperature 
phase is that of a half-metal, but with the substantial dip, 
almost a pseudogap at the Fermi level (see Ref.~\cite{sakamaki} 
for the result at $U=0$ eV).  Thus, one may anticipate that 
relatively minor changes in structure can push the occupied states 
down and empty states up, opening a gap in the spectrum and giving 
the gain in energy.  
This is indeed what happens in K$_2$Cr$_8$O$_{16}$: as shown in 
Fig.~\ref{fig2}(a)-(d), the calculations using the experimentally 
observed structure of the FI phase show that a gap of $0.03-0.13$ 
eV opens, depending on the value $U=2-3$ eV.  
This is in reasonable agreement with the gap $\sim$40 meV 
found recently by laser photoemission \cite{shin}.  

\begin{figure}[htbp]
\begin{center}
\includegraphics[scale=0.53]{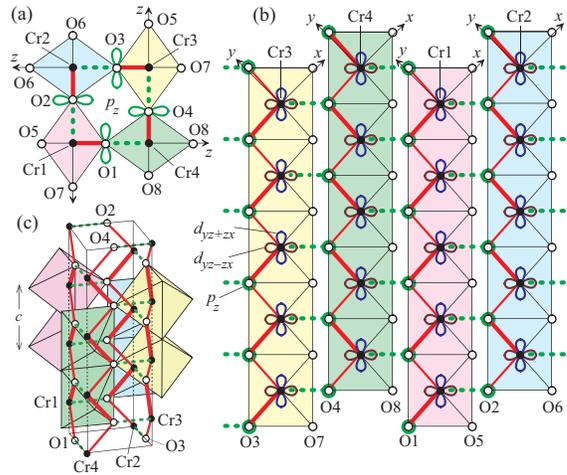}
\caption{(Color online) (a) The four-chain column viewed down 
the $c$-axis, (b) unfolded four-chain column, 
and (c) side view of the four-chain column.  
The shorter and longer bonds due to the lattice dimerization 
are indicated by thick and thin lines, respectively.  
Orbitals $d_{yz\pm zx}$ of Cr ions and $p_z$ of corner oxygens 
(O1, O2, O3, and O4) are also shown in the local coordinate axes 
$(x,y,z)$.}
\label{fig3}
\end{center}
\end{figure}

Our calculations confirmed that the ground state of K$_2$Cr$_8$O$_{16}$ 
is ferromagnetic both in the metallic and in the insulating phase.  
Calculations of the exchange coupling constants using the procedure 
of Ref.~\cite{LEIP} demonstrate that all the exchange integrals between 
nearest-neighbor Cr ions are indeed ferromagnetic, where, surprisingly, 
the strongest Cr-Cr coupling is not that in the double chains, but that 
{\it between} the double chains, {\it via} corner-shared oxygens 
(O1, O2, O3, and O4): in the metallic state, the values are $J\sim 5.3$ meV 
in the double chain, but twice as large between the double chains, 
$\sim$10.6 meV.  
The mechanism of ferromagnetism here is similar to that in CrO$_2$ 
\cite{korotin} and is in fact due to a double exchange \cite{sakamaki}.  
Indeed, one electron with the symmetry $d_{xy}$ in the local coordinate 
system of Fig.~\ref{fig3} (with the $z$-axis directed towards the 
corner oxygen connecting double chains) is practically localized, 
whereas the remaining electrons are in delocalized states formed by 
the $d_{yz}$ and $d_{zx}$ orbitals \cite{sakamaki,itoh}.  The opening of a 
small gap ($\sim$0.1 eV) at the Fermi level in the low-temperature 
phase, which is much smaller than the bandwidths $\sim$2 eV, does not 
significantly modify the exchange mechanism.  

The stronger coupling between the corner-sharing Cr ions of 
different double chains is clear from the orbital-projected density 
of states (DOS) shown in Fig.~\ref{fig2}(b)-(d); the states at the 
Fermi level are predominantly made up of the $d_{yz}$ and $d_{zx}$ 
orbitals of Cr with strong admixture of the $p_z$ orbitals of 
corner oxygens (O4) connecting two double chains, reflecting a 
negative charge-transfer-gap situation \cite{sakamaki}.  
Thus, it is natural to treat K$_2$Cr$_8$O$_{16}$ as consisting 
primarily of strongly coupled four corner-sharing Cr chains.  
In this sense, K$_2$Cr$_8$O$_{16}$ can in the first approximation 
be treated as a quasi-1D system, but with the 1D blocks quite 
different from what is usually assumed in the tunnel compounds of 
this type.  

The quasi-1D character of the electronic states is further 
confirmed by the Fermi surface calculated in the FM phase 
(see Fig.~\ref{fig2}(e)); the result also serves as the 
explanation of the nature of the MIT in this system.  
As one sees from Fig.~\ref{fig2}(e), there exists strong nesting 
in the electron spectrum, with the nesting wave vector $Q_z=2\pi/c$.  
Correspondingly, we may expect that there will occur a Peierls 
transition in this system.  The structural data presented above 
agree with this picture: they demonstrate that indeed a lattice 
dimerization in the $c$-direction corresponding to $Q_z$ occurs, 
after which the tetramers of Cr ions are formed in the four-chain 
columns.  This opens a gap at the Fermi level, but, as there are 
four Cr ions in a four-chain column per unit cell in the 
$c$-direction, it does not double the unit cell along $c$.  
On the other hand, the arrangement of these distorted four-chain 
columns leads to an increase of the unit cell in the $ab$-plane, 
with the unit cell becoming $\sqrt{2}\times\sqrt{2}\times 1$.  

The presence of 1D columns consisting of four Cr chains explains 
the existence of a Peierls transition for the average 
valence Cr$^{3.75+}$ in K$_2$Cr$_8$O$_{16}$.  
This value, with one extra electron per four Cr ions on top of 
the Cr$^{4+}$ background, seems at first glance to be unfavorable 
for Peierls dimerization.  However, for the 1D system of four chains, 
the spectrum for undimerized case has four fully spin-polarized bands 
near the Fermi level, touching at band edges (see Fig.~5 of 
Ref.~\cite{sakamaki} for the bands at $U=0$ eV), i.e., without a gap, 
but with the lowest band filled and 3 upper bands empty.  
The Peierls dimerization opens a gap between these bands.  

A tight-binding band calculation for an isolated column shown in 
Fig.~\ref{fig3} corroborates this picture; the calculation with the 
$d_{yz\pm zx}$ of Cr ions and $p_z$ orbitals of corner oxygens 
(O1, O2, O3, and O4), which are connected by $\pi$ bond, roughly 
reproduced the topmost four $t_{2g}$ bands of the GGA+$U$ band structure.  
Thus, the band structure near the Fermi level is indeed surprisingly 
simple.  The band gap opens in this tight-binding band structure when 
the Cr-O bond alternation observed experimentally in the FI phase 
was introduced.  Thus, the average valence Cr$^{3.75+}$ and the 1D 
columns made of four chains perfectly match, leading to a Peierls 
instability without any charge ordering.  

Finally, let us remark that the MIT due to Peierls distortion in 
a system of fully spin-polarized electrons, which leads to an 
uncommon FI phase of materials, is a very rare phenomenon.  
This effect is connected with the specific feature of the electronic 
structure of K$_2$Cr$_8$O$_{16}$, where the 1D objects in it are not 
the simple 1D chains but the four-chain columns.  
Usually, the Peierls distortion corresponds to a formation of 
valence bonds and requires the presence of both spin-up and 
spin-down electrons, whereas in the present case, we found that 
a similar effect can occur in a system of fully spin-polarized 
electrons.  In fact, it turned out that, in the present system, 
the electrons of only one spin close to the Fermi level determine 
the type of distortion and opening of the gap due to gain in kinetic 
(or band) energy via the Peierls mechanism, although the electron 
correlations among both-spin electrons are definitely important for 
preparing the state with a Peierls instability, i.e., the FM state 
of the double-exchange origin.  
We may therefore say that the present MIT is realization of 
a unique Peierls transition of fully spin-polarized electrons 
(or ``spinless'' fermions) where the spin degrees of freedom of 
electrons essentially play no role in the transition itself.  

In conclusion, we made the synchrotron X-ray diffraction experiment 
on the low-temperature FI phase of K$_2$Cr$_8$O$_{16}$ and reported 
the structural changes causing the MIT of this material.  
The structural distortions observed consist of dimerization in the 
four-chain columns of corner-sharing CrO$_6$ octahedra and formation 
of a stripe-like pattern of the dimerized columns.  These structural 
data, together with our electronic structure calculations, lead us to 
a model of the MIT in this system caused by the Peierls transition in 
the quasi-1D four-chain columns, which form the natural 1D building 
blocks.  
The 1D units in K$_2$Cr$_8$O$_{16}$ are thus different from the usually 
considered double chains.  The system remains ferromagnetic in the FI 
phase; the double-exchange ferromagnetism is not altered by the opening 
of the small gap.  
We thus demonstrated that the uncommon FI phase of materials can be 
realized in K$_2$Cr$_8$O$_{16}$ via the MIT due to Peierls mechanism 
in a system of fully spin-polarized electrons, which is indeed a very 
rare phenomenon.  
Besides specific importance for K$_2$Cr$_8$O$_{16}$, our results 
demonstrated that, in tunnel compounds like hollandites and similar 
materials, the coupling between the double chains via corner-shared 
oxygens may be as strong, and sometimes even stronger than that within 
these chains.  
This result may change the usual interpretation of the properties of 
the whole big class of these materials, which may have general and 
important implications for future researches.  

\begin{acknowledgments}
This work was supported by Kakenhi Grants 19052004, 21224008, 
22104011, 22244041, and 22540363 and FIRST program of Japan, by 
Russian program RFFI-19-02-96011, by German projects SFB 608, 
GR 1484/2-1, and FOR 1346, and by European project SOPRANO.  
This  study  was approved by PF-PAC 2009G200.  A part of 
computations was carried out at Research Center for Computational 
Science, Okazaki, Japan.  
\end{acknowledgments}

\end{document}